\documentclass{article}
\usepackage[final]{neurips_2024}
\usepackage{times}
\usepackage{titlesec}
\usepackage{xspace}
\usepackage{multirow} 
\usepackage{subcaption}
\usepackage{graphicx}
\usepackage{fdsymbol}
\usepackage{xcolor}
\usepackage{url}       
\usepackage{hyperref}  
\titleformat{\section}
  {\normalfont\Large\bfseries\MakeUppercase}{\thesection}{1em}{}

\definecolor{color1}{rgb}{0, 0.4470, 0.7410}
\definecolor{color2}{rgb}{0.9290, 0.6940, 0.1250}
\definecolor{color3}{rgb}{0.4660, 0.6740, 0.1880}
\definecolor{color4}{rgb}{0.6350, 0.0780, 0.1840}

\newcommand{\sysname}{TWSCardio\xspace}

\newcommand{\ie}{\textit{i}.\textit{e}.\xspace}
\newcommand{\eg}{\textit{e}.\textit{g}.,\xspace}
\newcommand{\etc}{\textit{etc}.\xspace}

\newcommand{\presec}{\vspace{-0.03in}}

\AtBeginDocument{%
  }

\title{Enabling Cardiac Monitoring using In-ear Ballistocardiogram on COTS Wireless Earbuds}

\author{%
  Yongjian Fu\thanks{ This work was conducted when the author was a visiting scholar at the University of California San Diego.} \  \thanks{Contributed equally to this work.} \\
  Central South University \\
  \texttt{fuyongjian@csu.edu.cn} \\
  \And
  Ke Sun\footnotemark[2] \\
  University of Michigan, Ann Arbor \\
  \texttt{kesuniot@umich.edu} \\
  \And
  Ruyao Wang \\
  University of Michigan, Ann Arbor \\
  \texttt{ruyao@umich.edu} \\
  \And
  Xinyi Li \\
  Tsinghua University \\
  \texttt{xinyili@tsinghua.edu.cn} \\
  \And
  Ju Ren\thanks{Corresponding author.} \\
  Tsinghua University \\
  \texttt{renju@tsinghua.edu.cn} \\
  \And
  Yaoxue Zhang \\
  Tsinghua University \\
  \texttt{zhangyx@tsinghua.edu.cn} \\
  \And
  Xinyu Zhang \\
  University of California, San Diego \\
  \texttt{xyzhang@ucsd.edu} \\
}

\begin{document}
\pdfoutput=1
\maketitle

\begin{abstract}
The human ear offers a unique opportunity for cardiac monitoring due to its physiological and practical advantages.
However, existing earable solutions require additional hardware and complex processing, posing challenges for commercial True Wireless Stereo (TWS) earbuds which are limited by their form factor and resources.
In this paper, we propose \sysname, a novel system that repurposes the IMU sensors in TWS earbuds for cardiac monitoring. 
Our key finding is that these sensors can capture in-ear ballistocardiogram (BCG) signals.
\sysname reuses the unstable Bluetooth channel to stream the IMU data to a smartphone for BCG processing.  
It incorporates a signal enhancement framework to address issues related to missing data and low sampling rate, while mitigating motion artifacts by fusing multi-axis information.
%
Furthermore, it employs a region-focused signal reconstruction method to translate the multi-axis in-ear BCG signals into fine-grained seismocardiogram (SCG) signals.
We have implemented \sysname as an efficient real-time app.
%
Our experiments on 100 subjects verify that \sysname can accurately reconstruct cardiac signals while showing resilience to motion artifacts, missing data, and low sampling rates.
Our case studies further demonstrate that \sysname can support diverse cardiac monitoring applications.
\end{abstract}

\vspace{-0.1in}
\section{Introduction}

Cardiovascular diseases (CVDs) are the leading cause of global mortality, accounting for approximately 17.9 million deaths annually \cite{di2024heart}. 
This staggering figure underscores the critical need for innovative solutions in cardiac care, particularly non-invasive, continuous monitoring technologies that can enhance early detection of CVDs while alleviating the burden on healthcare systems.
The widespread adoption of TWS earbuds, expected to reach over 700 million users by 2025 \cite{TWSmarket}, offers a promising mobile health platform to meet this demand.
%
The ears' anatomical structure, rich in vital blood vessels like the superficial temporal and posterior auricular arteries, provides a stable site for accurate cardiac monitoring. 

Current research in earable computing leverages these vascular advantages to develop cardiac monitoring solutions \cite{ferlini2021ear,fan2021headfi,fan2023apg,chen2024exploring}.
Representative techniques include integrating photoplethysmography (PPG) sensors for in-ear PPG monitoring \cite{ferlini2021ear} and using ultrasonic transducers to measure cardiac dynamics \cite{fan2023apg}.
Headphone drivers have also been repurposed \cite{fan2021headfi, chen2024exploring} to detect heartbeats and reconstruct phonocardiogram (PCG) signals.
However, these solutions necessitate additional hardware design and complex software processing. 
This poses significant challenges for commercial TWS earbuds, which are constrained by their form factor and resources, including energy, communication, and computation capabilities.

In this paper, we explore the potential of \textit{repurposing commercial TWS earbuds for cardiac monitoring applications.}
Our comprehensive preliminary studies demonstrate that standard IMU sensors, \ie, accelerometers and gyroscopes, embedded in TWS earbuds and originally intended for head motion tracking, can potentially capture in-ear BCG signals.
Based on this observation, we propose \sysname, a novel system that augments TWS earbuds with BCG monitoring capabilities. \sysname can convert the acquired BCG signals into SCG waveforms that are associated with clinically relevant cardiac metrics. 
It is designed to be compatible with COTS TWS earbuds, \eg Apple AirPods \cite{apple_airpods} and eSense \cite{kawsar2018earables}, and can be implemented through software enhancements, without any hardware modifications.
%
To maintain the earbuds' normal functions and minimize software overhead, \sysname utilizes the existing low-rate and unreliable Bluetooth Low Energy (BLE) communication channels to deliver raw IMU data to a mobile device (\eg, a smartphone), which then transforms the IMU data into BCG signals. 
\sysname employs an enhancement neural network to achieve resilience against unreliable sampling and user motion artifacts.
It further refines the low-rate and coarse-grained BCG signals into fine-grained and clinically reliable Seismocardiogram (SCG) signals by integrating data from multiple IMU axes and aggregating consecutive heartbeats.
Our case studies reveal that the recovered SCG signals can support various healthcare and HCI applications, including long-term Heart Rate (HR) and Heart Rate Variability (HRV) monitoring, biometric user identification, Blood Pressure (BP) estimation, and electrocardiogram (ECG) reconstruction.


To achieve these advantages, we address three  challenges:

\textit{Challenges of reusing the limited TWS earbuds hardware and software.} 
To preserve the earbuds' normal functions and minimize software overhead, \sysname utilizes the existing Bluetooth Low Energy (BLE) communication channel to stream raw IMU data to a mobile device (\eg a smartphone), which then processes the IMU data to derive BCG signals. 
However, the BLE-based IMU stream is primarily designed for real-time head motion tracking, which tends to be deprioritized by the TWS earbuds and allocated only minimal resources.   
%
Thus, the IMU data stream suffers from low sampling rate and unreliable sampling issues, especially when contending with more audio streaming functions. 
To overcome this challenge, we use the timestamps in the IMU stream to identify the locations of missing data.
We then develop a neural-based cardiographic continuity enhancement method by employing data fusion across multiple axes and ensemble learning from consecutive beats to fill in the missing data.
Additionally, we design a super-resolution scheme to reconstruct cardiographic signals at a high sampling rate.


\textit{Challenges in reliable in-ear BCG sensing.}
The BCG signals extracted from IMU typically exhibit low SNR compared to those from specialized sensors, especially in the presence of motion artifacts.
Additionally, variability in ear shapes and vessel locations results in differing sensitivities across the axes.
To address these challenges, we introduce a two-stage denoising solution. We first mitigate the motion artifacts in the frequency domain using stationary wavelet transform (SWT). Then we design a multi-head attenuation network to select and fuse the axes that are more sensitive to BCG signals.
%
%
Moreover, to enhance model generalization, 
we leverage an IMU simulator to create highly diverse training data, incorporating different motion artifacts and user profiles. 

\textit{Challenges of utilizing in-ear BCG for practical applications.}
The ultimate goal of \sysname is to harness in-ear BCG for practical applications, particularly in CVD monitoring.
However, existing neural-based signal enhancement solutions inadvertently suppress anomalous heartbeat signals. 
%
Similarly, anomalous heartbeat peak magnitude might be mistakenly normalized due to biases inherent in the training dataset.
To maintain the fidelity of crucial cardiac features, we carefully design our neural networks to achieve super-resolution BCG waveform reconstruction without changing the interval of the signals, thereby 
preserving essential time-domain information.
Additionally, we develop a cardiac-focused loss function that prioritizes the accurate reconstruction of peak regions of the cardiographic signals, while diminishing focus on less critical areas.
To further enhance accuracy, we design a transformer-based neural network that fuses and translates the multi-axis in-ear BCG signals into clinically reliable SCG signals \cite{wang2023knowing}, capitalizing on their shared waveform characteristics derived from mechanical heart movements.

We have implemented \sysname end-to-end as Android and iOS apps, running on smartphones  which acquire and process the IMU data from the TWS earbuds. 
We further collected a real-world dataset consisting of $40$ hours of in-ear BCG recordings from over $100$ participants, and synthesized an additional $150$ hours of data with $202$ types of motion artifacts for model training.
Our evaluation results show that \sysname effectively reconstruct SCG signals, with an average cosine similarity of $0.92$ across participants with varying BMI, ages, and genders.
\sysname estimates heart rate (HR) and inter-beat interval (IBI) with average errors of $2.73\%$ and $1.74\%$, respectively, even in the presence of significant motion artifacts, missing data, and low sampling rates.
\sysname operates in real-time on COTS TWS earbuds and paired smartphones with power consumption of only $1.14$ mAh/h and $7.26$ mAh/h, respectively.
Our case studies reveal that the recovered SCG signals can support various healthcare and HCI applications, including long-term HR and Heart Rate Variability (HRV) monitoring, biometric user authentication, pathological case detection, Blood Pressure (BP) estimation, and electrocardiogram (ECG) reconstruction.

We summarize our contributions as follows:

$\bullet$ We conduct comprehensive preliminary studies and demonstrate that standard IMU sensors embedded in TWS earbuds can effectively capture in-ear BCG signals.

$\bullet$ We develop an end-to-end, software-only BCG sensing scheme that can be seamlessly integrated into COTS TWS earbuds without requiring any hardware modifications while minimizing software overhead.

$\bullet$ We design a neural enhancement network to mitigate the motion artifacts and enhance cardiographic signal continuity.  We further devise a region-focused signal reconstruction method that fuses and translates multi-axis, coarse-grained in-ear BCG signals into fine-grained SCG signals.  

$\bullet$ We conduct comprehensive experiments to demonstrate \sysname's accuracy, reliability, and effectiveness in supporting diverse use cases.

\section{Background and Related Work}
This section first introduces the representative cardiographic signals measured by wearable devices.
Then, we compare existing earable sensing solutions with \sysname.

\subsection{Earable Cardiographic Signal Sensing}
Emerging TWS earbuds are not only equipped with advanced acoustic features like active noise cancellation (ANC) but also increasingly used for mobile health. For example, Apple has incorporated hearing aid functionalities that allow users, particularly the elderly, to personalize their sound profiles based on hearing capabilities \cite{Apple_HearingAid}. This demographic trend, coupled with the devices' technological advancements, amplifies their suitability for ubiquitous health monitoring. 
%

\noindent \textbf{In-ear PPG}: 
Research has incorporated PPG sensors in TWS earbuds
\cite{ferlini2021ear}.
These sensors facilitate respiratory \cite{romero2024optibreathe} and blood pressure monitoring by analyzing variations between the left and right earbuds \cite{balaji2023stereo}. 
However, PPG technology requires additional sensors, increasing hardware complexity and limiting its feasibility for large-scale deployment.
Audioplethysmography (APG) \cite{fan2023apg} transmits low-intensity ultrasound signals into the ear and captures reflected echoes, producing signals with waveforms similar to PPG signals.
Although promising, APG's complex processing of high-frequency ultrasound signals adds computational load to earbuds, while limited Bluetooth bandwidth constrains raw data transmission, leaving its compatibility with commercial TWS earbuds uncertain.
%

\noindent \textbf{In-ear PCG}:
The ear canal's natural acoustics amplify heart sounds, facilitating in-ear PCG measurements \cite{cao2023heartprint}.
Utilizing in-ear microphones, studies have captured these sounds for biometric verification \cite{cao2023heartprint} and blood pressure estimation \cite{zhao2024hearbp}.
However, the computing and communication resources of standard TWS earbuds are insufficient to accommodate the additional overhead associated with high-frequency in-ear microphone signal processing and transmission.
Moreover, ambient noise remains a significant hurdle for in-ear PCG, even with in-ear microphone setups.
Alternatively, HeadFi \cite{fan2021headfi} and Asclepius \cite{chen2024exploring} repurpose headphone drivers as sensors to detect heartbeats and reconstruct PCG signals.
These approaches, however, involve intricate hardware additions like low-noise amplifiers and voltage converters, which increase the complexity, cost, and energy consumption of the earbuds. 

\noindent \textbf{Around-ear BCG:}
The work most closely related to \sysname uses IMU sensors around the ear to measure heartbeat signals. FaceReader \cite{zhang2023passive, zhang2023facereader} explores the potential of using IMU sensors in an AR/VR headset for heartbeat detection. Since these sensors are positioned far from the arteries, they primarily detect bone vibrations caused by heartbeats, resulting in heart-rate estimation only.
%
Conversely, \cite{winokur2012wearable} integrates IMU sensors into a custom-designed hearing aid, firmly attaching them behind the ear to capture \textit{behind-ear} BCG signals. 

\textit{To our knowledge, \sysname represents the first system to use the existing IMU sensors in TWS earbuds to measure in-ear BCG signals and reconstruct SCG/ECG signals. This method is implementable on COTS TWS earbuds without any hardware modifications, changes to wireless communication protocols, or significant system overhead.}

\begin{figure}[!t]
\setlength{\abovecaptionskip}{2pt}
\setlength{\belowcaptionskip}{-10pt}
\centering
\begin{minipage}[t]{0.4\linewidth}
\centering
\includegraphics[width=0.75\linewidth]{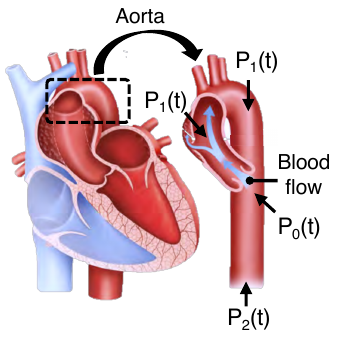}
\caption{The causes and model of BCG formation.}
\label{fig:bcgModel}
\end{minipage}
\hspace{1mm}
\begin{minipage}[t]{0.4\linewidth}
 \centering
\includegraphics[width=0.99\linewidth]{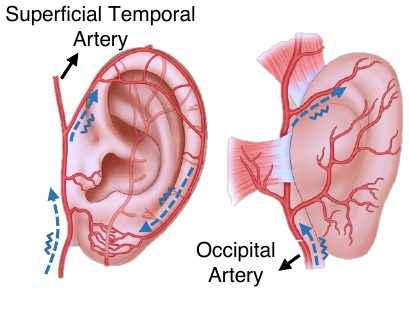}
 \caption{BCG propagation near the ears.}
 \label{fig:bcgEar}
\end{minipage}
\end{figure}

\section{Preliminary Study}
\label{sec:prelim}

Figure \ref{fig:bcgModel} shows the basic in-ear BCG model, which can be interpreted as the blood-pressure gradients in the ascending and descending aorta \cite{kim2016ballistocardiogram}.
When an instantaneous force is exerted on the blood in the main artery, the BCG force $F_{BCG}$ can be estimated as

\vspace{-4.5mm}
\begin{small}
\begin{equation}
\label{bcg_model}
    F_{BCG}(t) = S_d[P_1(t) - P_2(t)] - S_a[P_0(t) - P_1(t)]
\end{equation}
\end{small}
\vspace{-4.5mm}

\noindent where $S_d$ and $S_a$ are the average cross-sectional areas of the descending and ascending aorta, respectively. $P_0(t)$, $P_1(t)$ and $P_2(t)$ represent the blood pressure waves of the ascending aorta inlet, arch outlet/inlet, and descending aorta outlet, respectively.
%
As shown in Figure \ref{fig:bcgEar}, two arteries around the ear, \ie superficial temporal artery and occipital artery, introduce significant mechanical vibration inside the ear.
\textit{Our key finding is that in-ear BCG signals can propagate into the ear canal, allowing them to be measured by the IMU sensors in COTS TWS earbuds.}

\subsection{Feasibility of In-ear BCG}\label{sec:feasibility}
\begin{figure}[!t]
\setlength{\abovecaptionskip}{2pt}
\setlength{\belowcaptionskip}{-2pt}
\centering
    \begin{subfigure}{0.45\textwidth}
        \includegraphics[width=\linewidth]{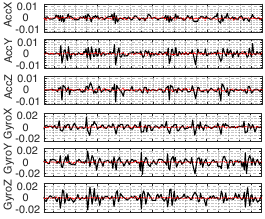}
        \caption{Left-ear BCG from \#A}
        \label{fig:BCG_SubjectA_left}
    \end{subfigure}
    \begin{subfigure}{0.45\textwidth}
        \includegraphics[width=\linewidth]{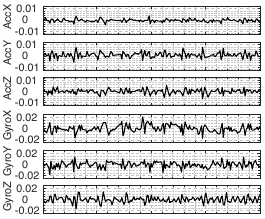}
        \caption{Right-ear BCG from \#B}
        \label{fig:BCG_SubjectB_left}
    \end{subfigure}
    \caption{Raw IMU waveform collected by Apple AirPods Pro (Duration: 8 seconds; Sampling Rate: 25 Hz; Units: accelerometer $m/s^2$ and gyroscope $dps$). Red dotted line represents the noise floor when the AirPods Pro is placed on a flat surface.}
    \label{fig:raw_waveform}
\end{figure}

To assess the feasibility of in-ear BCG, we conducted preliminary studies using two commercial TWS earbuds that provide access to raw accelerometer and gyroscope data from their IMU sensors, \ie the Apple AirPods series \cite{apple_airpods} and the eSense open earable platform \cite{kawsar2018earables}. 
%
During the experiment, subjects were instructed to wear the earbuds as they normally would while remaining stationary.
Figure~\ref{fig:raw_waveform} shows the raw IMU waveform recorded by Apple AirPods Pro (2nd generation), which exhibits the same pattern to typical BCG signals \cite{kim2016ballistocardiogram}. 
%
We repeated this experiment using eSense and found that it also consistently captured in-ear BCG signals.
%
This attributes to the high resolution of existing IMUs.
According to the datasheet, earable accelerometers have a typical measurement range of $\pm2 g$ and a digital resolution of 16 bits, translating to a resolution of $0.06 mg/LSB$. The gyroscopes feature a measurement range of $\pm125 dps$ (degrees per second) and a digital resolution of 16 bits, resulting in a resolution of $0.004 dps/LSB$. 
Our preliminary study indicates that the average maximum peak amplitude of in-ear BCG, characterized by accelerometer and gyroscope, is approximately $0.005 m/s^2$ and $0.03 dps$, respectively, resulting in an average empirical peak SNR of $38 dB$ and $19 dB$.
\textit{This suggests that the IMU sensors in TWS earbuds have sufficient resolution to sense in-ear BCG signals.}

\begin{figure}[!t]
\setlength{\abovecaptionskip}{2pt}
\setlength{\belowcaptionskip}{-10pt}
\centering
    \includegraphics[width=0.7\linewidth]{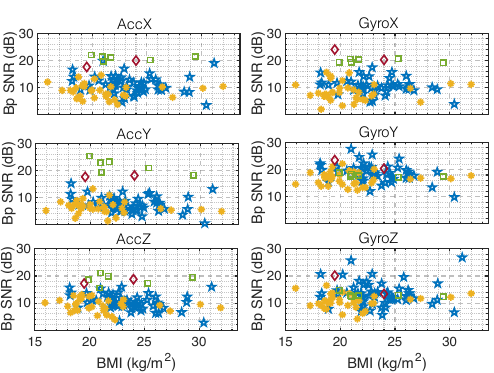}
    \caption{Frequency domain Bandpass SNR benchmark across different IMU axes and subjects. Each marker represents a specific subject. ``\textbf{\textcolor{color1}{$\medwhitestar$}}'' and ``\textcolor{color2}{$\ast$}'' denote male and female subjects using Apple AirPods, respectively, while ``\textcolor{color3}{$\medsquare$}'' and ``\textcolor{color4}{$\meddiamond$}'' indicate male and female subjects using eSense, respectively. All figures share the same x-axis representing the subject's BMI.}
    \label{fig:pre_demo_SNR}
\end{figure}

To measure the reliability of in-ear BCG, we employ the frequency domain Bandpass SNR \cite{mckay1970signal}, 
which evaluates how effectively the $1\sim10Hz$ cardiac frequency components are preserved amidst noise.
%
The Bandpass SNR (in dB) is calculated as: 
$10 \log_{10}(P_{signal}/P_{noise})$, 
where we compute the power spectral density (PSD) to determine the mean power of both the signal ($P_{signal}$) and noise ($P_{noise}$).  
$P_{noise}$ is sampled when placing the earbuds on a stationary surface. 

We measured the SNR of multi-axis in-ear BCG signals across $100$ subjects. Sec.~\ref{sec:implementation} elaborates on our participants' profiles. 
%
Our findings are illustrated in Figure~\ref{fig:pre_demo_SNR} and summarized as follows.

\noindent \textbf{Variable SNR across different IMU axes.}
We observe significant variation in SNR across different IMU axes.
The axis oriented toward the occipital artery (Y-axis for Apple AirPods and X-axis for eSense) achieves the highest SNR, averaging $18.21 dB$ and $20.51 dB$, respectively.
Conversely, the axis facing outward from the subject (X-axis for Apple AirPods and Z-axis for eSense) exhibit the lowest SNRs, with averages of $10.04 dB$ and $13.93 dB$, respectively.

\noindent \textbf{Consistent SNR across subjects of different gender, BMI, and age.}
Female subjects exhibit slightly lower SNRs than male subjects, with an average difference of $1.53 dB$ across different axes.
BMI affects the SNR, but insignificantly, with only a $1.87 dB$ reduction observed between underweight (BMI $< 18.5 kg/m^2$) and obese subjects (BMI $> 30 kg/m^2$).
With respect to age, the average SNR is $12.33 dB$ for subjects under 30 years of age, $10.67 dB$ for those between 30 and 40 years, and $9.56 dB$ for those over 40 years.
This decrease in SNR with age is likely due to the higher BMI typically associated with older subjects.

\noindent \textbf{Consistent SNR across different ear tips and wearing styles.}
The SNR is consistent for different ear tip size and wearing styles.
Apple provides four ear tip sizes and a fit test to select the best size for users \cite{ear_tip_fit_test}.
When users chose an unsuitable ear tip without passing the ear tip fit test, the SNR drops by only $1.34 dB$ and $1.13 dB$ for the accelerometer and gyroscope, respectively.  

\noindent \textbf{Consistent SNR across different COTS TWS earbuds.}
The IMU sensors in AirPods are not configurable, whereas eSense provides IMU sensor configuration.
We configure the eSense to maximize SNR, by minimizing the measurement range of accelerometer with $\pm2g$ and gyroscope with $\pm250dps$, respectively, while disabling the on-chip low-pass filter. 
eSense achieves higher SNR than AirPods Pro, with an average SNR of $18.84 dB$ compared to $13.12 dB$.
We suspect that this is also because eSense can be better secured on the subject's ears as it offers specific designs to ensure a firm attachment \cite{eSense_doc}.

%




\subsection{Challenges of In-ear BCG}\label{sec:challenges}
Despite the potential of in-ear BCG sensing, several key challenges must be addressed for practical usage.


\begin{figure}[!t]
\setlength{\abovecaptionskip}{2pt}
\setlength{\belowcaptionskip}{-10pt}
\centering
 \includegraphics[width=0.8\linewidth]{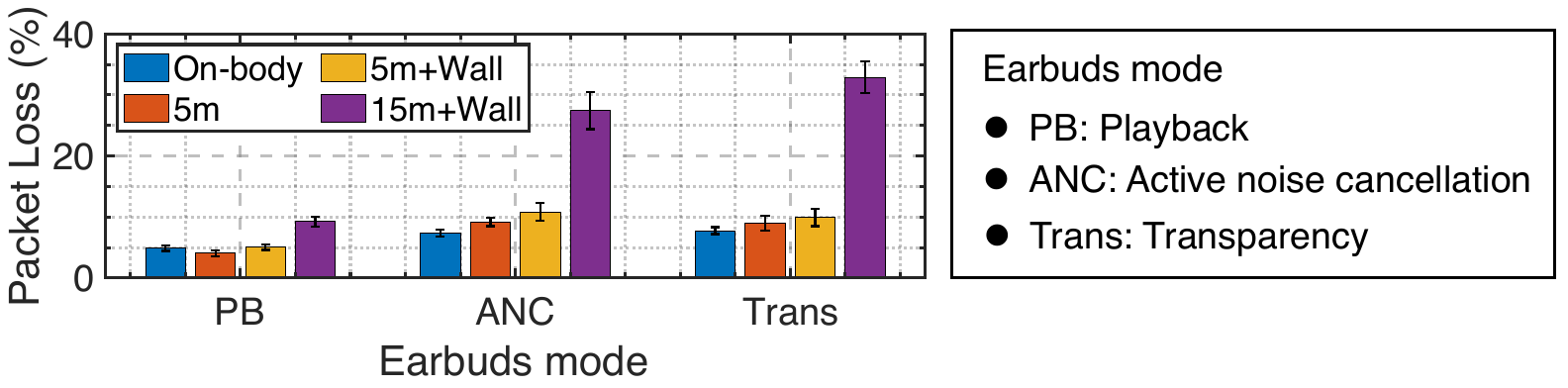}
  \caption{IMU stream BLE channel packet loss rate.}
  \label{fig:pre_packetloss}
\end{figure}

\subsubsection{Inherent Challenges due to Earbuds' Hardware 
 and Software Constraints.}
Because of the limited computational resources, the earbuds' IMU data have to be streamed to an associated mobile device for post processing, which involves two challenges. 

\noindent \textbf{Low Sampling Rate.} Standard IMU sensors typically support sampling rates exceeding $500 Hz$. However, TWS earbuds are limited to much lower rates due to BLE constraints.
COTS TWS earbuds limit each BLE packet to transfer one 6-axis IMU data sample without buffering to guarantee real-time processing.
%
The minimal packet interval for Bluetooth 5.4 is $7.5 ms$, theoretically allowing a maximum rate of $133.3 Hz$.
%
%
In practice, factors such as energy consumption and computational limitations further reduce this rate.
For example, Apple AirPods Series is configured with a minimal packet interval of $40 ms$, resulting in a $20\sim25 Hz$ rate.
%
These rates are sufficient for head motion tracking, but inadequate for cardiac monitoring and analysis, 
which typically demands a minimum sampling rate of $100\sim200Hz$ \cite{sadek2019ballistocardiogram}.

\noindent \textbf{Unreliable Sampling.}
The IMU data stream transmitted over BLE exhibits inconsistent sampling intervals.
We examined the IMU stream from the Airpods Pro as received by an iPhone.
Based on the received timestamps, we found that the IMU data is uniformly sampled at $40~ms$ intervals with a $25~Hz$ sampling rate. However, \textit{the sampling intervals deviate significantly, by multiples of $40~ms$, indicating data loss}.
Figure \ref{fig:pre_packetloss} shows the measured timestamp skips in the IMU stream under different scenarios.
Significant data loss occurs due to several factors.
First, the computational resources allocated to other real-time audio functions impact the IMU sampling reliability. 
IMU sample loss is particularly severe during ``ANC'' and ``Trans'' modes.
Second, data loss is also influenced by BLE channel conditions.
When the smartphone is within $5$ meters of the subject or on their body, packet loss is minimal. However, with $15$~m separation and wall obstructions, packet loss exceeds $30\%$.
Besides, the real-time control algorithms for head motion tracking and spatial sound generation in TWS earbuds necessitate a short supervision timeout (\ie, $100 ms$) on BLE channels to ensure low latency.
The supervision timeout is the maximum period a BLE device can wait before considering the connection lost. 
A short supervision timeout exacerbates data loss, especially in a busy wireless channel. 
This observation aligns with other measurement studies of BLE wearable devices \cite{tipparaju2021mitigation}.

\begin{figure}[h]
\setlength{\abovecaptionskip}{2pt}
\setlength{\belowcaptionskip}{-10pt}
\centering
 \includegraphics[width=0.8\linewidth]{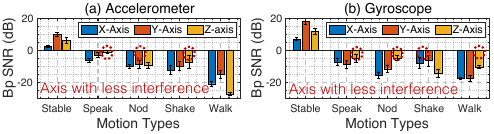}
  \caption{In-ear BCG SNR under common motions.}
  \label{fig:motion_SNR}
\end{figure}

\subsubsection{Challenges in Practical Usage Scenarios.}
\hspace{1em}\\
\textbf{Low SNR}.
We compared the SNR of in-ear BCG with that of other body parts.
During the measurement, users were instructed to firmly attach an AirPod earbud to the ``Back of the ear'', ``Chest'', and ``Neck'' using tape. 
%
In-ear BCG has $7.97dB$, $6.76dB$ and $2.27dB$ lower SNR than ``Back of the ear'', ``Chest'', and ``Neck'', respectively.
This is because the two near-ear arteries
are not proximate to the in-ear measurement site (Figure~\ref{fig:bcgEar}).
Besides, 
in-ear BCG were captured with normally worn earbuds, rather than the tape-secured setting. 

\noindent \textbf{Motion artifacts}. 
As illustrated in Figure~\ref{fig:motion_SNR}, we quantified the SNR of in-ear BCG across several common motions, including ``Speak'', ``Nod'', ``Head shake'', and ``Walk''. ``Speak'' motion introduces interference at intensities comparable to that of stable in-ear BCG. In contrast, head movements and walking produce substantial motion noise, surpassing the desired BCG signals by several orders of magnitude.

\noindent \textbf{Variation across different subjects}.
Although different subjects achieve similar SNR, their in-ear BCG waveforms exhibit significant variability.
Figure~\ref{fig:raw_waveform} demonstrates an example for two subjects.
Subject \#A's in-ear BCG is sensitive to the Y-axis and Z-axis of the accelerometer and the Y-axis and Z-axis of the gyroscope, while Subject \#B is sensitive only to the X-axis and Y-axis of the gyroscope.
This variability is attributed to differences in ear shapes, vessel locations, and wearing styles among subjects, which affect the waveform and the sensitivity of different IMU axes.

\section{System Design}
\label{sec:design}
\begin{figure*}[!t]
\setlength{\abovecaptionskip}{2pt}
\setlength{\belowcaptionskip}{-10pt}
\centering
 \includegraphics[width=0.85\linewidth]{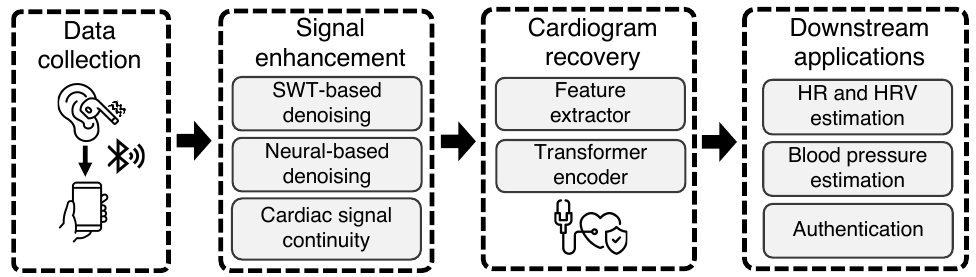}
  \caption{System overview.}
  \label{fig:overview}
\end{figure*}

In this section, we introduce the key design components of \sysname which follow the workflow in Figure~\ref{fig:overview}. 

\subsection{Motion Artifacts Denoising}\label{sec:motion_denoising}

Although motion artifacts obscure the subtle in-ear BCG in the time-frequency domain, as visualized in Figure~\ref{fig:mtd_denoise}(b), two key factors offer new opportunities to denoise such interference.
First, in-ear BCG has a unique cardiac frequency pattern distinct from arbitrary body motion.
This pattern enables us to determine whether the IMU signals are dominated by motion artifacts or in-ear BCG, allowing for targeted enhancement of the relevant signals.
Second, certain sensors and axes among the 6-axis IMU experience less motion interference during specific motions, as highlighted in the example in 
Figure~\ref{fig:motion_SNR}.
Our motion artifact denoising pipeline leverages these two findings, comprising two stages: SWT-based denoising and neural-based multi-axis denoising.

\subsubsection{SWT-based denoising.}
Common motion artifacts, particularly those generated by slow head movements and systemic body motions, typically have frequency components below $2~Hz$ \cite{nagai2017motion}.
In contrast, the critical peaks of BCG signals, including the \textit{H, I, J, K} and \textit{L} peaks, occur at frequencies above $3 Hz$ \cite{ochoa2022signal}.
\begin{figure}[h]
\setlength{\abovecaptionskip}{2pt}
\setlength{\belowcaptionskip}{-10pt}
\centering
 \includegraphics[width=0.8\linewidth]{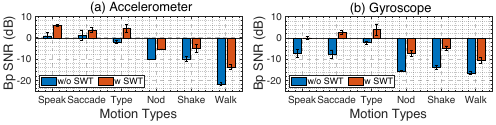}
  \caption{SNR w/ and w/o SWT denoising.}
  \label{fig:swt_denoise}
\vspace{-0.05in}
\end{figure}
Our SWT-based algorithm is designed to reduce the low-frequency dominant motion interference while maintaining the critical high-frequency BCG signals.

\begin{figure*}[t]
\centering
\includegraphics[width=0.98\textwidth]{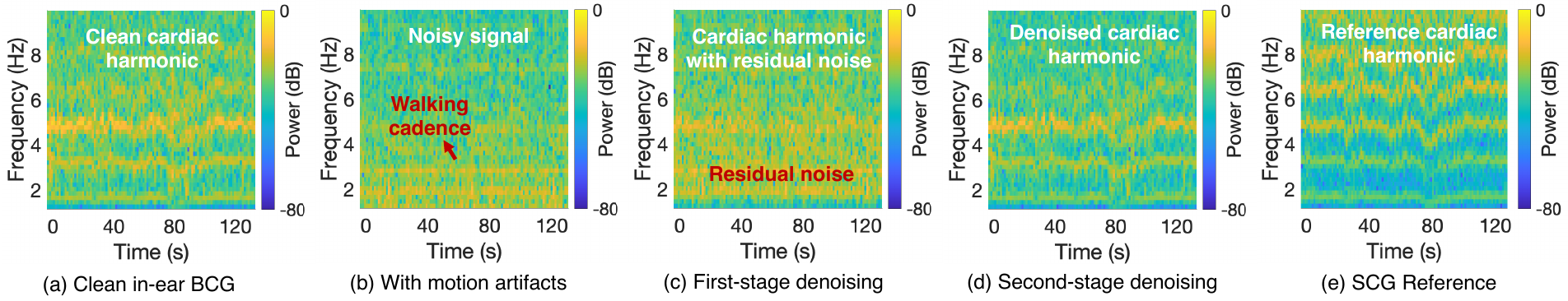}
    \caption{Time-frequency domain visualization for motion artifacts denoising.}
\label{fig:mtd_denoise}
\end{figure*}

The SWT-based denoising method can be expressed as
\begin{small}
\vspace{-2mm}
\begin{equation}
    s(t) = \sum_{i=1}^{J} D_i^{b}(t) + \sum_{i=J + 1}^{N} D_i^{m}(t)
    \vspace{-2mm}
\end{equation}
\end{small}
where $D_i^{b}$ and $D_i^{m}(t)$ represent the $i$th wavelet coefficients corresponding to high-frequency BCG signals and lower-frequency motion artifacts, respectively.
$N$ is the order of SWT, and $J$ is the decomposition level separating the frequencies of motion artifacts and BCG signals.
We set $N = 5$ and $J = 2$ when the sampling rate of IMU sensor is $f_s = 25$ Hz.
Therefore, the filtered signal frequency band is $> f_s/2^{J+1} = 3.25$ Hz, ensuring the preservation of critical peak information in BCG signals.
In practice, we first decompose the IMU signals using SWT, and then reconstruct the enhanced time-domain cardiac signals $s_e(t)$ by applying inverse SWT to high-frequency component $\{D_1^{b}, D_2^{b}\}$.
Note that these steps are performed separately for each axis.

%
%
%
%
%
%

Figure~\ref{fig:swt_denoise} shows the SNR with and without SWT-based denoising under various motion artifacts.
SWT-based denoising effectively reduces low-frequency motion interference, notably during subtle, slow motions like ``Speak'', ``Saccade'' and ``Type''.
However, significant residual noise remains in axes with high-frequency interference, mostly due to fast motions like ``Nod'', ``Shake'', and ``Walk'' (see Figure~\ref{fig:mtd_denoise}(c)).

\subsubsection{Neural-based multi-axis denoising.}\label{sec:neural_denosing}
To reduce the residual noise, 
we design a denoising neural network that takes the SWT-denoised IMU data as input, and fuses the multi-axis data based on the aforementioned two findings. 

\noindent \textbf{Model design.} Figure~\ref{fig:mtd_deepDenoiseNet} illustrates our model architecture.
The model processes 5-second, 6-axis IMU data sampled at 25 Hz, leveraging a channel attention module to enhance SNR and a denoising autoencoder with an encoder-decoder structure to extract cardiac signals. The encoder compresses the input to filter noise, while the decoder reconstructs clean signals by focusing on cardiac components and suppressing motion artifacts.
%
%
First, we design a channel attention module that uses multi-head attention to dynamically leverage the channels and time periods with higher SNR to denoise the low-SNR channels.
Second, we incorporate a denoising autoencoder, which has proven effective in denoising time-series signals and reconstructing complex signal patterns \cite{lu2013speech}.
%
%

\noindent \textbf{Training data synthesis.} 
Training a robust denoising model is challenging due to the difficulty of obtaining paired BCG data with and without motion artifacts. To overcome this, we utilize the additive nature of motion artifacts and synthesize training data by combining clean BCG signals with simulated motion signals, enabling effective model training \cite{seok2021motion}.

%
%
%

To collect clean in-ear BCG signals, we reuse the \textit{100 subjects}, instructing them to remain seated and stationary.
The resulting data include more than $145,000$ beats, totaling around 25~hours. 
As for the motion artifact, direct collection of IMU data will inadvertently mix the motion and BCG signals. 
Additionally, a massive training dataset is needed to cover a wide variety of motion types and ensure model generalization.
Fortunately, existing 
research has developed IMU simulators capable of accurately converting human body motion data into accelerometer and gyroscope signals. 
We adapt the IMU simulator in \cite{huang2018deep} to synthesize motion artifact-induced earbud IMU signals.
Specifically, we use the AMASS database \cite{mahmood2019amass}, a large collection of human motion data that unifies 15 optical marker-based motion capture datasets.
It provides $63$ hours of high-quality, high-resolution 3D human meshes in Skinned Multi-Person Linear (SMPL) format, covering a wide variety of daily human motions, including locomotion, exercising, cooking, and human-object interaction, \etc
We synthesize the training IMU data by linearly combining the self-collected in-ear BCG signals with the simulated motion artifact signals.
%
%
%
Our final synthesized dataset contains $150$ hours of data, over $300$ different body shapes, and a wide variety of daily motions. 
This comprehensive dataset facilitates the training of a generalizable denoising neural network.  

\subsection{Cardiographic Continuity Refinement} \label{sec:continuity}

\begin{figure}[h]
\setlength{\abovecaptionskip}{2pt}
\setlength{\belowcaptionskip}{-10pt}
\centering
 \includegraphics[width=0.6\linewidth]{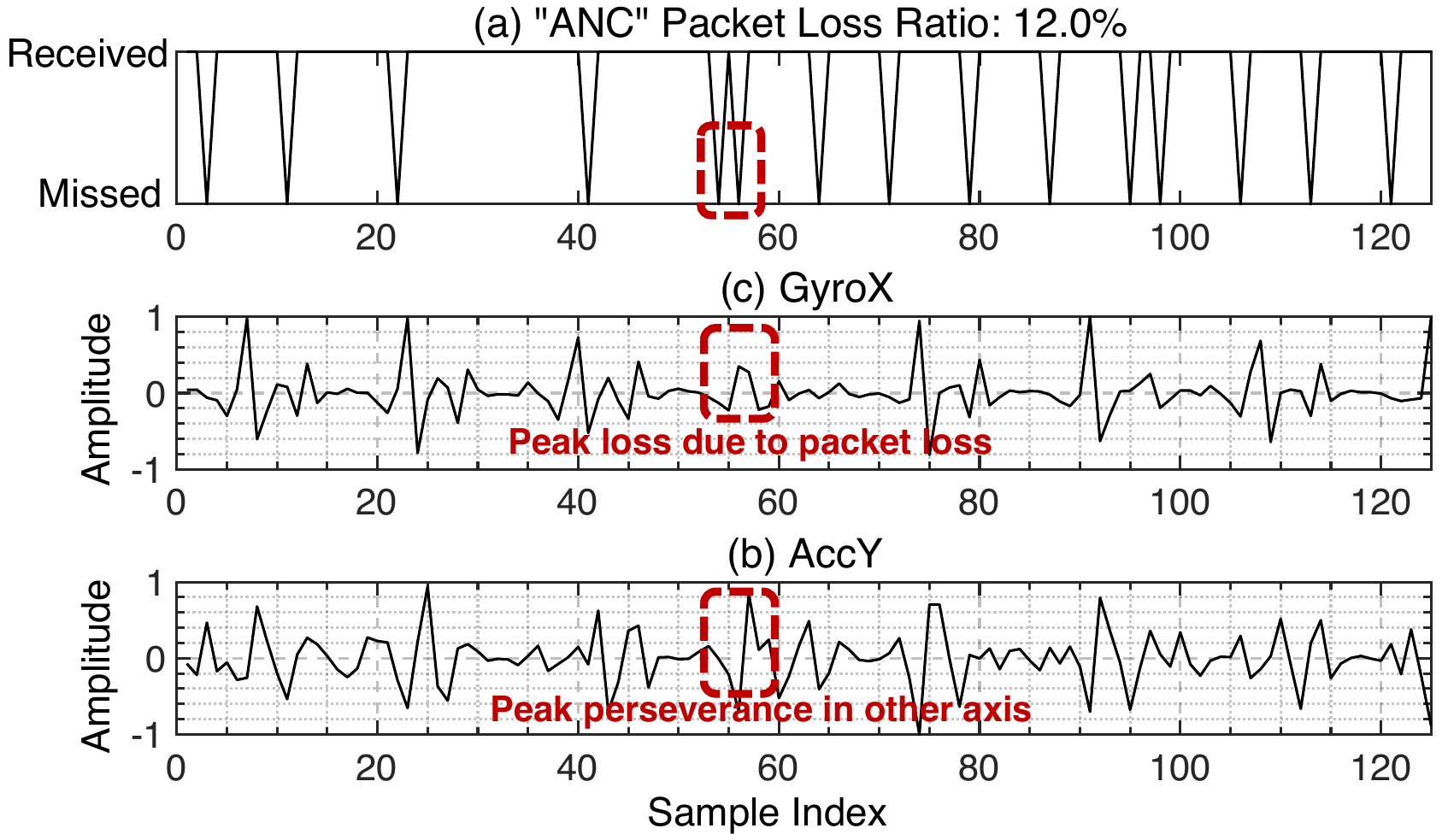}
  \caption{Example of missing data issue.}
  \label{fig:mtd_packetloss}
\end{figure}

Our signal enhancement method aims to further address the challenges of cardiographic discontinuity caused by unreliable sampling and data loss.
Figure~\ref{fig:mtd_packetloss} shows a real-world example of an IMU stream received by a smartphone when a stationary subject wore the earbuds in ANC mode.
Even after applying linear interpolation, the in-ear BCG signals remain significantly disrupted, particularly when critical peak information is lost, as highlighted in Figure~\ref{fig:mtd_packetloss}(b).

\begin{figure}[h]
\setlength{\abovecaptionskip}{2pt}
\setlength{\belowcaptionskip}{-10pt}
\centering
 \includegraphics[width=0.6\linewidth]{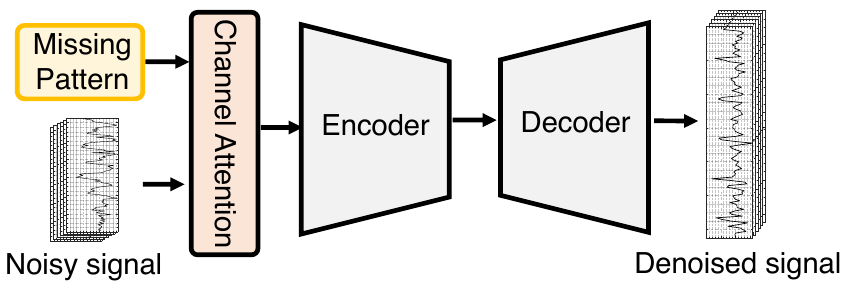}
  \caption{Deep denoising model for motion artifacts.}
  \label{fig:mtd_deepDenoiseNet}
\end{figure}

\sysname leverages two principles to design the cardiographic continuity refinement solution.
First, the peak positions in BCG signals vary across different axes due to the transmission of vibrations through blood vessels in multiple directions \cite{inan2014ballistocardiography}.
Therefore, even if a specific axis (Figure~\ref{fig:mtd_packetloss}(b)) misses critical peak information, other axes can preserve their own (Figure~\ref{fig:mtd_packetloss}(c)).
%
Second, consecutive beats exhibit similar BCG signals and are unlikely to miss the same peak information simultaneously.
Thus, they can be combined to resolve cardiographic discontinuities.
These two design principles align closely with the neural-based multi-axis denoising model.
Consequently, the same neural network in Section~\ref{sec:neural_denosing} can be repurposed to achieve dual objectives: motion artifact denoising and cardiographic continuity refinement.
To guide the model to focus on missing data refinement, the corresponding pattern of missing samples, extracted from IMU stream timestamps, is integrated into the model.
As shown in Figure \ref{fig:mtd_deepDenoiseNet}, \sysname employs the missing pattern as a binary mask, providing supplementary positional information for guiding the model to recover the missing data.

%
To train this model, we collected clean in-ear BCG signals, and
%
%
%
%
then randomly generate a missing pattern with varying packet loss rates (0\% to 50\%) and apply it to the clean data.

\noindent \textbf{Detecting the limit of cardiographic continuity refinement.}
Our cardiographic continuity refinement solution may not perform well when the packet loss rate is significantly high, \eg, due to a long BLE link distance. 
To address this, we set a packet loss rate threshold $\tau$ to identify when the IMU stream is too severely corrupted. \sysname can accordingly enter a fail-safe mode, warning the user to discard such data. 
We empirically determine $\tau = 24\%$ by evaluating the performance of our cardiographic continuity refinement under different packet loss rate, as detailed in. 
\subsection{Cardiogram Reconstruction}\label{sec:cardiogram_recovery}

Although our signal enhancement design improves the in-ear BCG quality, its low sampling rate and unreliable waveform limit its ability to provide detailed cardiac insights. 
Whereas the IMUs on COTS TWS earbuds can achieve $40~ms$ time resolution (at $25~Hz$), reliable HRV measurement requires a time resolution of at least $10~ms$ \cite{johnston2020heart}.
Besides, multi-axis in-ear BCG provides inconsistent peak magnitude and waveform information, which varies depending on the subject's ear shape, vessel locations, and wearing style. 
%
To overcome the challenge, we develop a super-resolution cardiogram reconstruction neural network that translates multi-axis in-ear BCG into clinically reliable SCG signals with a high sampling rate.

\begin{figure}[!t]
\setlength{\abovecaptionskip}{2pt}
\setlength{\belowcaptionskip}{-10pt}
\centering
 \includegraphics[width=0.7\linewidth]{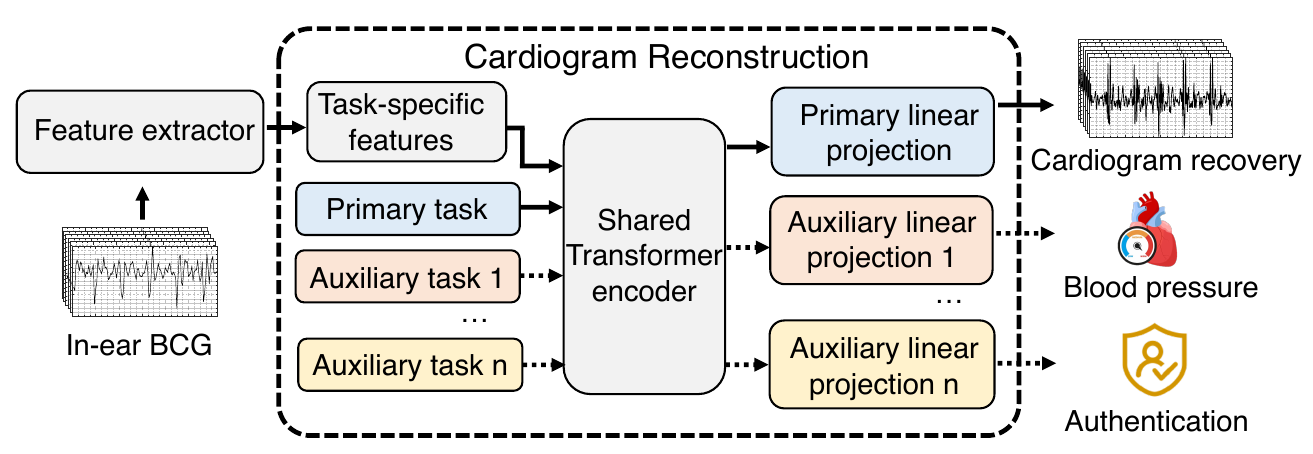}
  \caption{Cardiogram Reconstruction.}
  \label{fig:mtd_cardiogramRecovery}
\end{figure}

\noindent \textbf{Model design.} Our model input is the output from the aforementioned denoising neural network. 
%
%
%
%
%
As shown in Figure~\ref{fig:mtd_cardiogramRecovery}, our model adopts a systematic approach to process in-ear BCG signals, starting with a feature extraction stage that identifies and preserves key spatial and temporal characteristics. This is followed by a pattern recognition module that captures long-range dependencies and highlights critical signal features related to heartbeat dynamics. The final stage transforms these processed features into refined output signals by leveraging Transformer~\cite{fu2022svoice}, ensuring high fidelity and consistency with the underlying physiological patterns. Throughout the process, the model is designed to balance complexity and efficiency, allowing it to perform well under various real-world conditions. 
The model is trained using our self-collected 25~hours of clean in-ear BCG data and their corresponding SCG ground truth (Section~\ref{sec:implementation}).

\noindent \textbf{Adapation to various applications using foundational cardiac-related embeddings.}
In-ear BCG signals can be used not only to reconstruct the SCG signals for measuring HR and HRV but also for other diverse applications (Sec.~\ref{sec:casestudy}). 
%
However, each application requires collecting a large amount of paired in-ear BCG data and corresponding ground truth to train a specific model. 
To overcome this burden, we leverage the features learned from the aforementioned BCG-to-SCG super-resolution translation task as \textit{foundational embeddings}. 
Instead of retraining an entire model, we only need to replace the final fully connected layer and fine-tune the model for the specific task using few-shot learning.
Our results in Section~\ref{sec:casestudy} show that we need less than $5\%$ of the data to achieve reasonable performance for various applications, compared to training the original BCG-to-SCG reconstruction model.

\section{Implementation}\label{sec:implementation}

\begin{figure}[h]
\setlength{\abovecaptionskip}{2pt}
\setlength{\belowcaptionskip}{-10pt}
\centering
 \includegraphics[width=0.4\linewidth]{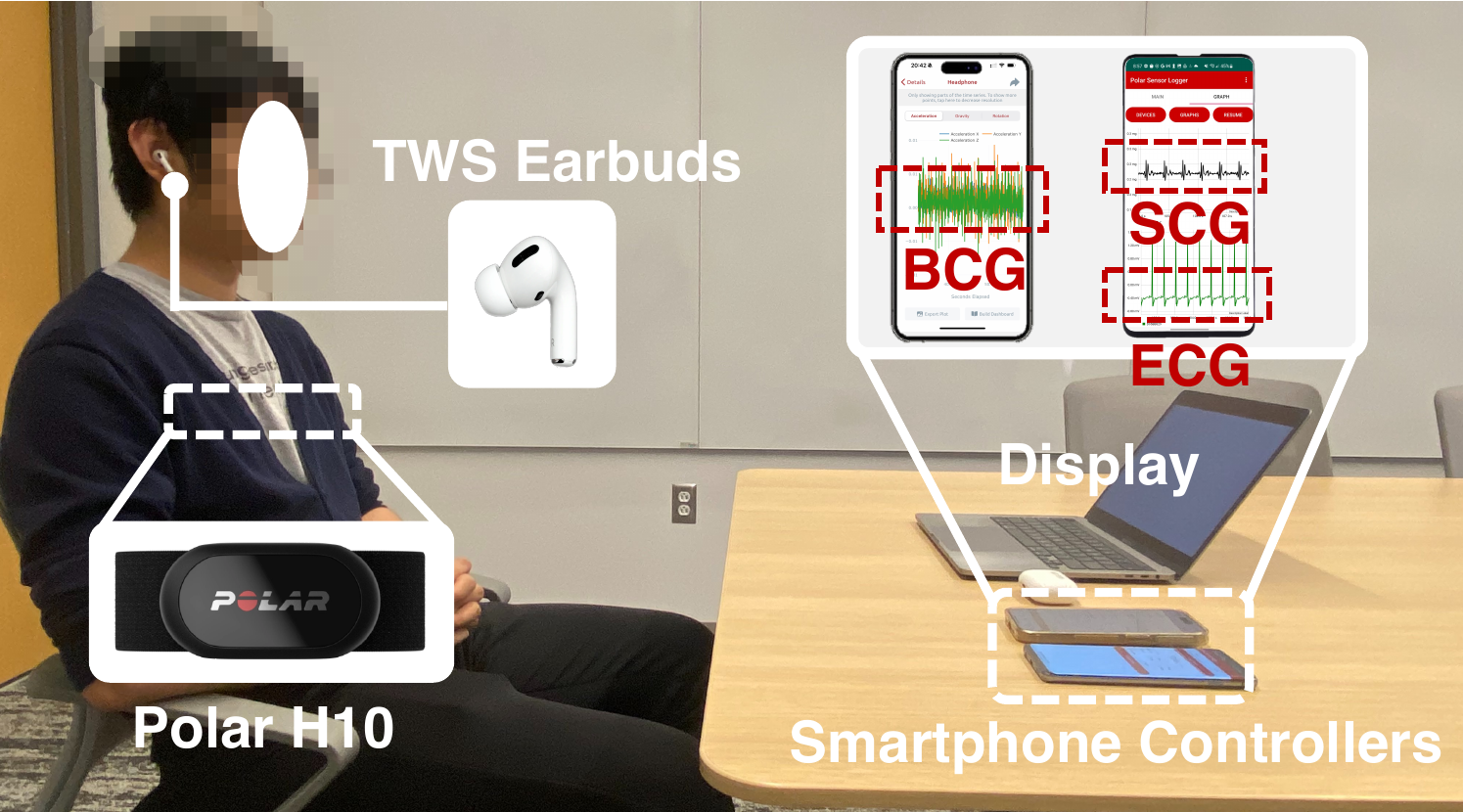}
  \caption{Experimental setup.}
  \label{fig:setup}
\end{figure}

\textbf{Data collection.} As depicted in Figure~\ref{fig:setup}, our experimental setup consists of a TWS earbud paired with a smartphone.
We implement apps to acquire 6-axis raw IMU data using native iOS API  \cite{apple_CMHeadphoneMotionManager} and eSense Android library \cite{eSense_android} for AirPods and eSense, respectively.
Since these two achieves comparable performance (Sec.~\ref{sec:prelim}), our experiments use AirPods by default, unless noted otherwise.  
%

To collect clean in-ear BCG data in a controlled setting, participants were seated in a chair and instructed to wear the right earbud normally by default.
%
The Polar H10 was used for ground truth according to the user manual \cite{polor10_usermanual}.
%
%
%
A total of $100$ participants took part in the experiments, including $40$ females and $60$ males, aged $19$ to $62$ years, with body mass indices (BMI) ranging from $15.92 kg/m^2$ to $33.74 kg/m^2$.
%
%

%
%
%
%
The detailed testing process will be discussed in each experiment.
SCG signals were not considered as ground truth data in the testing dataset due to significant interference from motion artifacts.
Instead, we use Polar H10 to collect $130Hz$ ECG signals and precise heartbeat rate (HR), inter-beat interval (IBI) measurements calculated from ECG signals as the ground truth.
%
All experimental procedures were approved by an IRB.

\noindent \textbf{Model training.}
%
Both signal enhancement network (Sections~\ref{sec:motion_denoising} and \ref{sec:continuity}) and cardiogram reconstruction model (Section~\ref{sec:cardiogram_recovery}) were trained on a server equipped with an A40 GPU. 
%
%
The dataset was divided into $10$ groups for cross-validation, with each fold involving training on $90$ participants and validation on the remaining $10$.

\noindent \textbf{Pipeline implementation on smartphone.}
We implemented the end-to-end pipeline of \sysname (Figure~\ref{fig:overview}), including data capturing, signal processing and neural network model inference, as a single iOS APP on the iPhone 15 Pro running iOS 18.
Neural network inference was implemented using Core ML, provided by Apple Developer \cite{apple_coreml}.
With Core ML, we were able to switch the computational unit for neural network inference between the CPU, GPU, and Neural Engine, for system performance evaluation.


\section{Evaluation}
\begin{figure*}[t]
\centering
\includegraphics[width=1.0\textwidth]{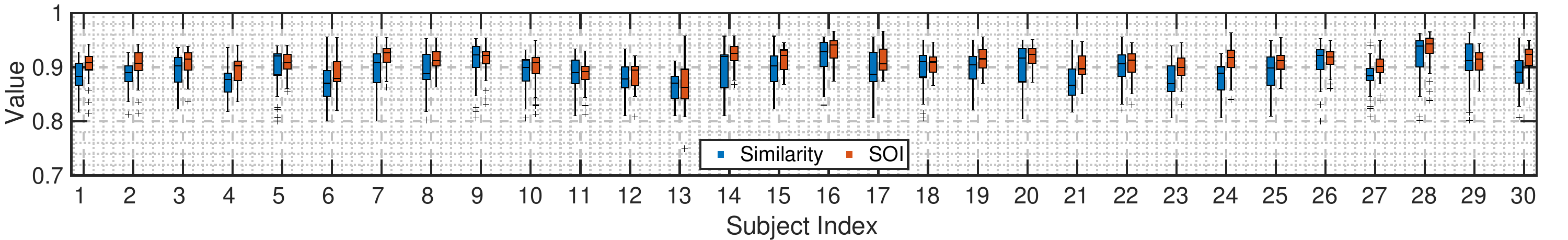}
    \caption{Overall performance across 30 random selected subjects.}
\label{fig:exp_overall}
\end{figure*}

\subsection{Evaluation Metrics}
We use three metrics to assess the quality of the enhanced in-ear BCG and reconstructed cardiographic signals. 

\noindent $\bullet$ \textit{Cosine Similarity (Similarity)} evaluates how closely the reconstructed signal aligns with the ground truth in the time domain, with values closer to 1 indicating higher similarity.\\
\noindent $\bullet$ \textit{Spectral Overlap Index (SOI)} measures frequency-domain accuracy, with values near 1 indicating strong spectral fidelity. 
It is calculated by comparing the PSD between the recovered waveform \(\text{PSD}_{\text{e}}(f)\) and the ground truth \(\text{PSD}_{\text{g}}(f)\) within the frequency range \([1Hz, 50Hz]\), defined as:

\vspace{-3mm}
\begin{small}
\begin{equation}
\text{SOI} = {\sum_{f=1Hz}^{50Hz} \min\left(\text{PSD}_{\text{e}}(f), \text{PSD}_{\text{t}}(f)\right)}\Big/{\sum_{f=1Hz}^{50Hz} \text{PSD}_{\text{t}}(f)}
\end{equation}
\end{small}
\vspace{-3mm}

\noindent $\bullet$ \textit{Mean Percentage Error (MPE)} quantifies the percentage error in HR and IBI estimates, two widely used parameters for CVD monitoring \cite{nelson2019accuracy}, by comparing the estimated values \(V_{e,i}\) with the true values \(V_{t,i}\) across \(N\) measurements.

\vspace{-3mm}
\begin{small}
\begin{equation}
\text{MPE} = \frac{1}{N} \sum_{i=1}^{N} \left| \frac{V_{e,i} - V_{t,i}}{V_{t,i}} \right| \times 100\%
\end{equation}
\end{small}
\vspace{-3mm}

Similarity and SOI are used to evaluate SCG reconstruction, which require the ground truth SCG as reference.
Similarity and SOI values above $0.9$ and $0.85$ are considered indicative of a \textit{very high} and \textit{high} positive correlation for SCG reconstruction, respectively \cite{ha2020contactless, mukaka2012guide}.
MPE is used to assess the accuracy of HR and IBI measurements, with ground truth data derived from ECG signals.
MPE values below $5\%$ are used as a threshold for good accuracy in HR and IBI measurements in wearable or mobile health device studies \cite{nelson2019accuracy}.
\begin{figure}[h]
 \centering
\includegraphics[width=0.4\linewidth]{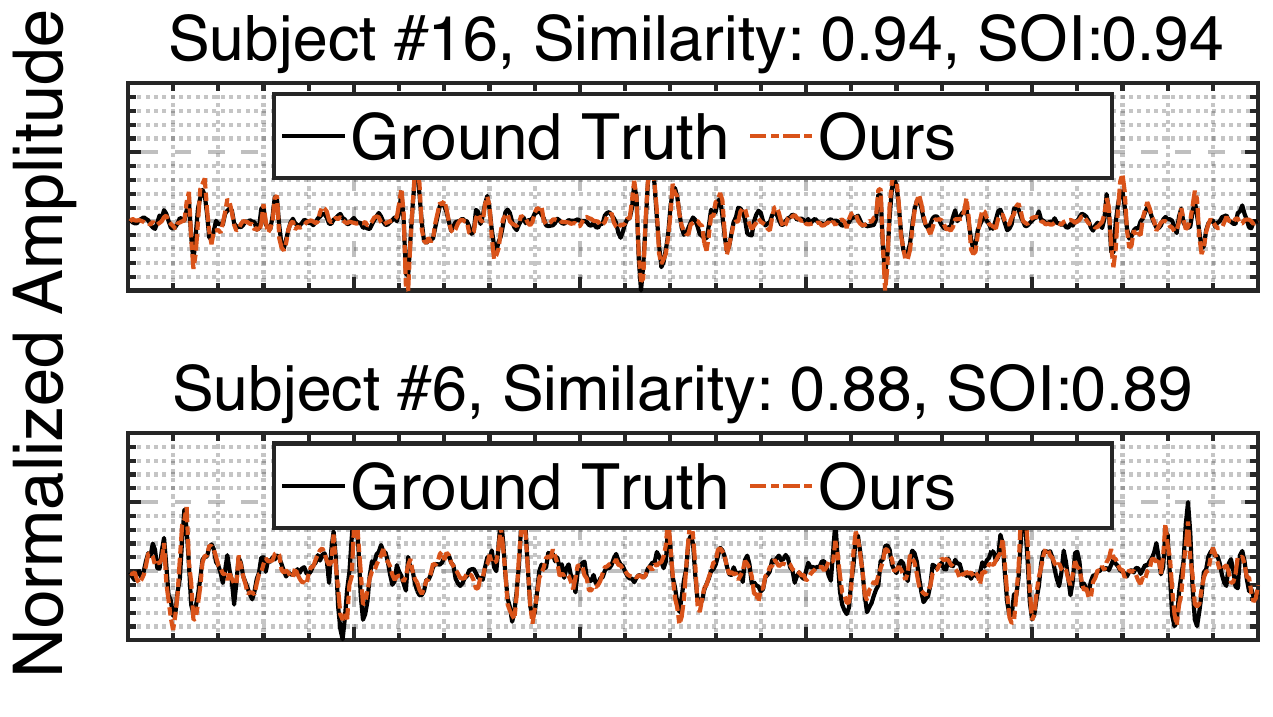}
 \caption{Reconstructed SCG.}
 \label{fig:exp_scgRecSample}
\end{figure}

\subsection{End-to-end Performance}
%
We first evaluate the end-to-end SCG reconstruction performance to show the generalization of \sysname across different scenarios.
%
For testing, $212$ types of motion artifacts IMU data are generated using the IMU simulator, while missing data patterns were captured in real-world scenarios under various earbud operating modes. Both are then applied to the real-world in-ear BCG testing dataset.
The results are reported using $10$-fold cross-subject validation.

\begin{figure}[h]
 \centering
\includegraphics[width=0.35\linewidth]{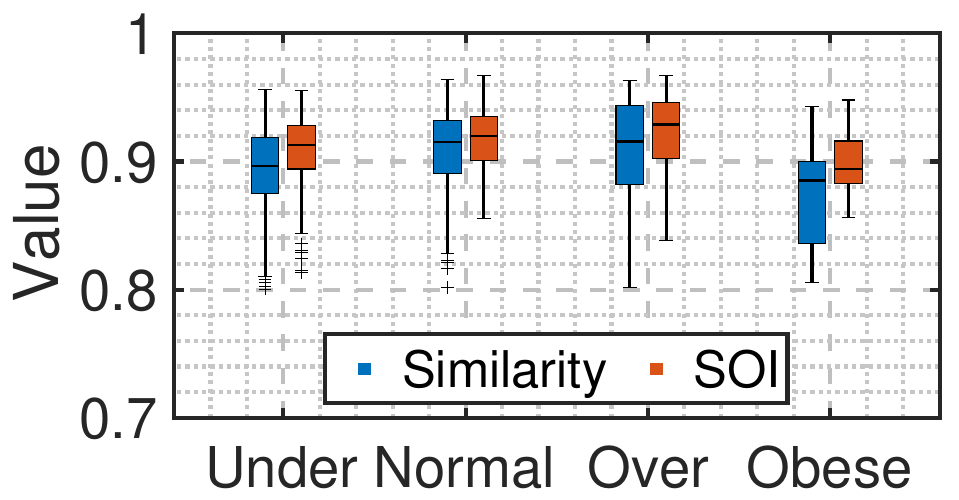}
 \caption{Impact of BMI.}
 \label{fig:exp_bmi}
\end{figure}

\sysname demonstrates a \textit{very high} positive correlation for SCG reconstruction, achieving an average Similarity of 0.92 and an SOI of 0.93 across 100 subjects.
Due to space limits, we randomly select $30$ participants and visualize the results in Figure~\ref{fig:exp_overall}, and shown few reconstruction smaples in Figure~\ref{fig:exp_scgRecSample}.
These results demonstrate \textit{strong generalization of \sysname across different subjects}.

\noindent \textbf{Impact of BMI.}
We divide the 100 subjects into four groups based on BMI: ``Under'' ($<18.5 kg/m^2$), ``Normal'' ($18.5-24.9 kg/m^2$), ``Over'' ($25.0-29.9 kg/m^2$), and ``Obese'' ($>30.0 kg/m^2$). As shown in Figure~\ref{fig:exp_bmi}, only the obese group deviates slightly from other groups, but even this group achieves a high accuracy (Similarity: $0.87$, SOI: $0.90$). This suggests that 
\textit{BMI has minimal impact on \sysname}.

\begin{figure}[h]
 \centering
\includegraphics[width=0.35\linewidth]{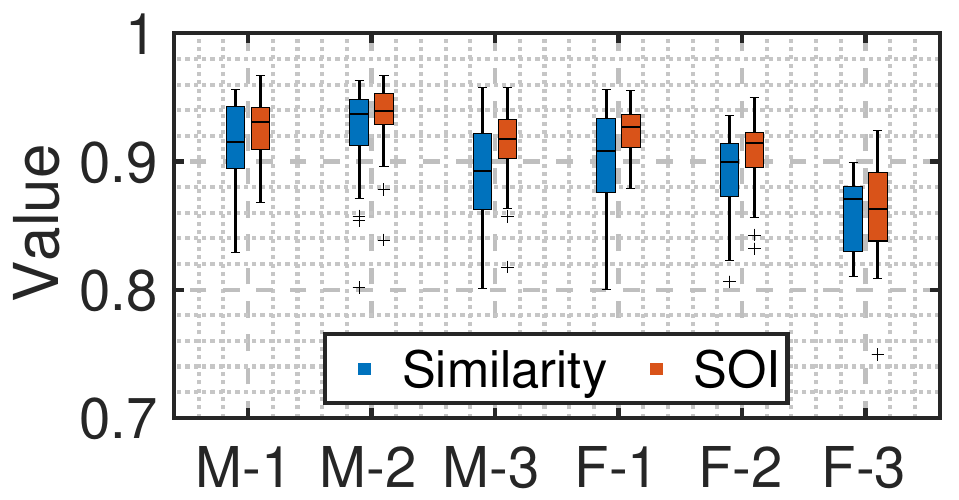}
 \caption{Age and gender.}
 \label{fig:exp_genderAge}
\end{figure}

\noindent \textbf{Impact of age and gender.} We categorize the $100$ subjects into six groups based on gender (``F'' for female, ``M'' for male) and age (``1'' for $<26$ years, ``2''for $26$–$45$ years, ``3'' for $>45$ years). As shown in Figure~\ref{fig:exp_genderAge}, \textit{\sysname performs consistently across both age and gender}, with average Similarity and SOI values exceeding $0.9$ in most groups.
However, in the F-3 group, both Similarity and SOI fall slightly below $0.9$. We suspect that \textit{age-related changes, such as reduced cardiac force, arterial stiffening, and muscle atrophy, likely contribute to the observed reduction in Similarity and SOI}~\cite{sadek2019ballistocardiogram}. We plan to include more representative samples from this group in future studies to better understand this observation.

\begin{figure}[h]
 \centering
\includegraphics[width=0.35\linewidth]{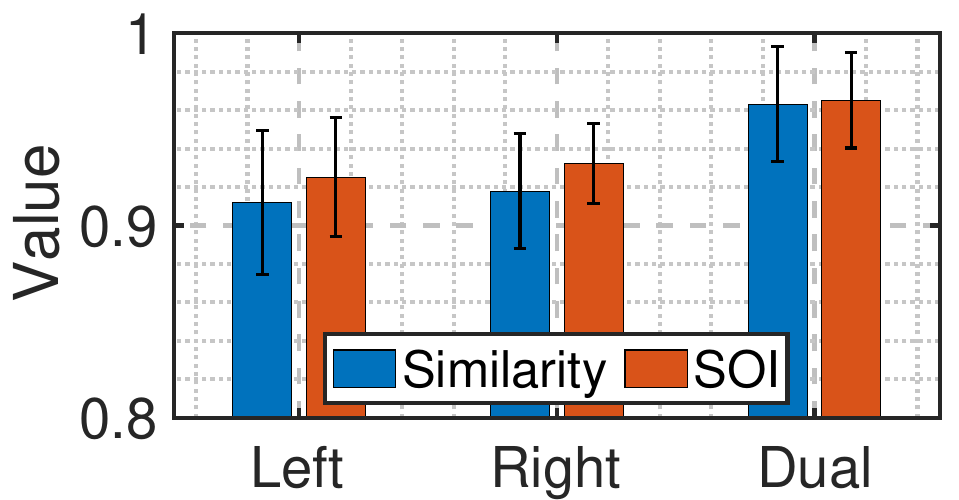}
 \caption{Side of earbuds.}
 \label{fig:exp_numEarbuds}
\end{figure}

\noindent \textbf{Impact of earbuds side.}
Our default model is trained using right-ear data.
To explore the effect of earbuds side, we invite an additional 10 participants to use dual-ear setup for data collection.
We fine-tune the network's CNN and linear projection layers to train models for left-ear and dual-ear setup while freezing the Transformer parameters to preserve its sequence modeling capabilities.
Figure~\ref{fig:exp_numEarbuds} shows the results for the these three setups. The right-ear model performs slightly better than the left, likely due to the model's inherent bias from being predominantly trained on right-ear data.
We expect that the left-ear model can improve by balancing the training data from both ears.
When utilizing dual-ear setup, the Similarity and SOI both elevate to 0.96, demonstrating that the bilateral in-ear BCG signals offer substantial complementary information. 
This finding strongly supports the \textit{use of dual-ear IMU data streams for fine-grained cardiac sensing}.

\subsection{Case Study} \label{sec:casestudy}
We conducted $4$ case studies to show the usage of \sysname.
%
%
%
%
%
%

\subsubsection{Pathological Case Detection.} To demonstrate that \sysname can preserve anomalous heart conditions, we recruit three patients known with anomaly heart conditions: tachycardia, bradycardia, and arrhythmia, corresponding to resting heart rate exceeding 100~BPM, below 60~BPM, and varying HR and HRV, respectively.

\begin{figure}[h]
 \centering
\includegraphics[width=0.55\linewidth]{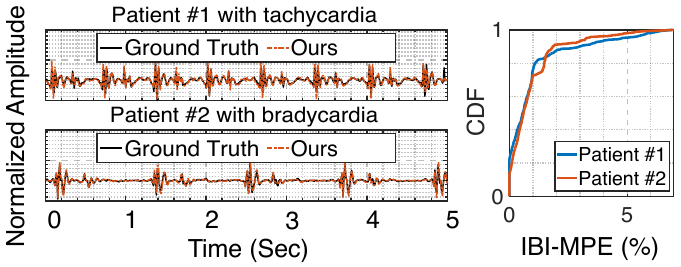}
 \caption{Analysis of pathological cases.}
 \label{fig:exp_abnormal}
\end{figure}
Figure~\ref{fig:exp_abnormal} presents reconstructed SCG waveform for the first two cases, demonstrating that \sysname achieves robust IBI estimation, with MPE below $2\%$ in $90\%$ of cases, underscoring its \textit{effectiveness in accurately capturing anomaly cardiac signals}.
%
%
This suggests that \textit{\sysname can identify the pathological cases for early-stage CVD detection.}
Our future plan is to evaluate \sysname across a broader cohort of CVD patients. 


\subsubsection{User authentication.} We leverage \sysname for biometric user authentication, which can be used to enhance device security while enabling personalized TWS earbud experiences.
A $5\%$ of in-ear BCG data from $50$ participants in our default dataset was used to fine-tuned the pre-trained cardiogram reconstruction model (Section~\ref{sec:cardiogram_recovery}) into an authentication model by adjusting the linear projection layers.
We use the Area Under the Curve (AUC), a common metric for user authentication \cite{seok2023photoplethysmogram}.
\begin{figure}[h]
 \centering
\includegraphics[width=0.35\linewidth]{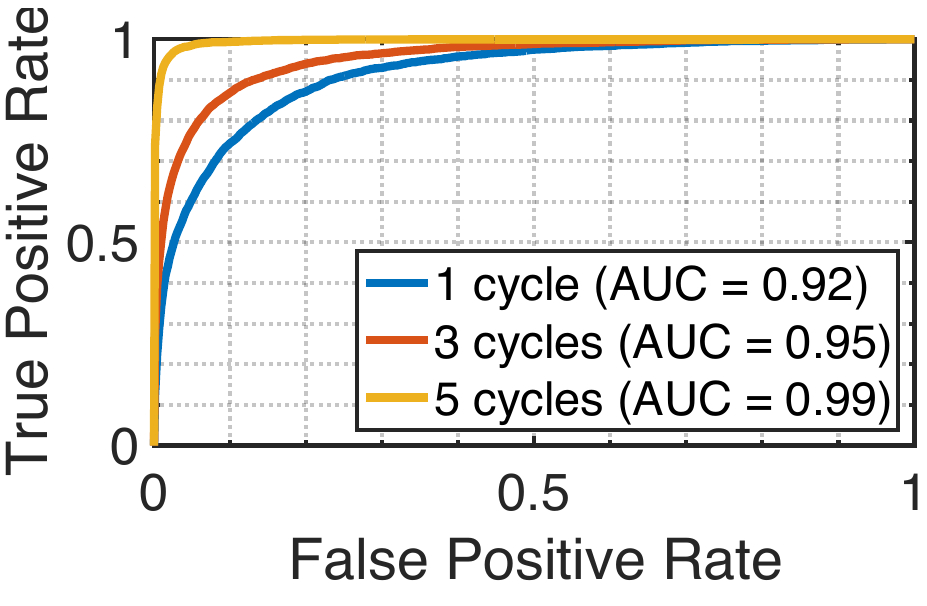}
 \caption{Authentication.}
 \label{fig:exp_authentication}
\end{figure}
An AUC $>0.9$ is considered excellent for biometric authentication \cite{seok2023photoplethysmogram}.
As shown in Figure~\ref{fig:exp_authentication}, \sysname achieves AUCs of $0.92$, $0.95$ and $0.99$ with $1$, $3$ and $5$ heartbeat cycles as model input, respectively.
This highlights the possibility of \textit{using \sysname for quick and reliable earbud-based user authentication}.

\begin{figure}[h]
 \centering
\includegraphics[width=0.55\linewidth]{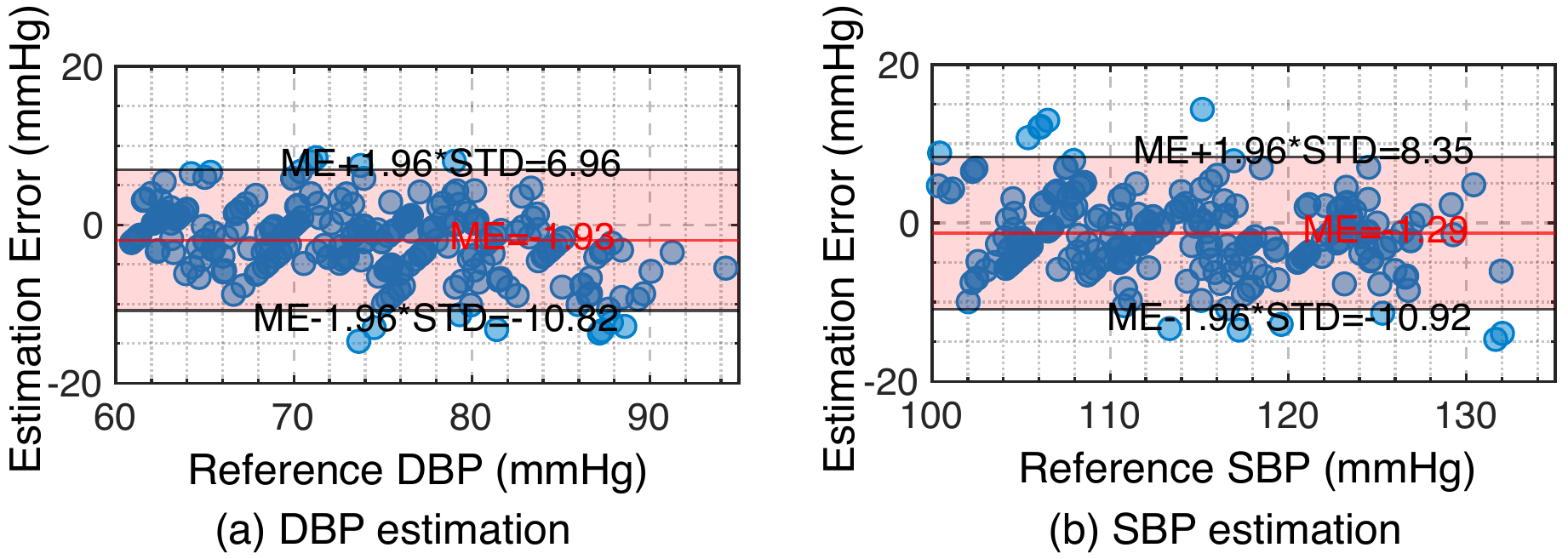}
 \caption{Blood pressure estimation.}
 \label{fig:exp_bp}
\end{figure}
\subsubsection{Blood pressure estimation.}
%
Previous study shows that BP can be derived by analyzing the time differences in cardiac signal propagation between multiple measurement points, \ie, Pulse Transit Time (PTT)~\cite{balaji2023stereo}.
Inspired by this, we propose to leverage the dual-ear channels as natural reference points to capture PTT for BP estimation. 
We recruited $5$ participants to collect an additional $6$-hour dual-ear in-ear BCG data, while measuring the ground truth BP using an FDA-certified Checkme BP2A Monitor over one week.
We also fine tune the pre-trained cardiogram reconstruction network for BP estimation, using $4$ hours of data for training and $2$ hours for testing.
\sysname achieves a correlation of 0.81 for Diastolic Blood Pressure (DBP) and 0.83 for Systolic Blood Pressure (SBP). 
Figure~\ref{fig:exp_bp} shows that the mean error for DBP and SBP is $1.93mmHg$ and $1.29 mmHg$, meeting the clinical accuracy standards for BP estimation.
Our future work will test the BP estimation model on a larger number of participants.
\subsubsection{Electrocardiogram reconstruction.}
We further leverage \sysname to reconstruct the in-ear BCG into ECG, the gold standard for CVD monitoring.
We collected an additional $2.5$-hour dataset from $10$ participants over one week using a dual-ear setup along with Polor H10 ECG signals as the ground truth.
The model was fine-tuned from the pre-trained cardiogram reconstruction network using a $2$-hour training dataset and evaluated on $30$-min testing dataset.
\begin{figure}[h]
 \centering
\includegraphics[width=0.55\linewidth]{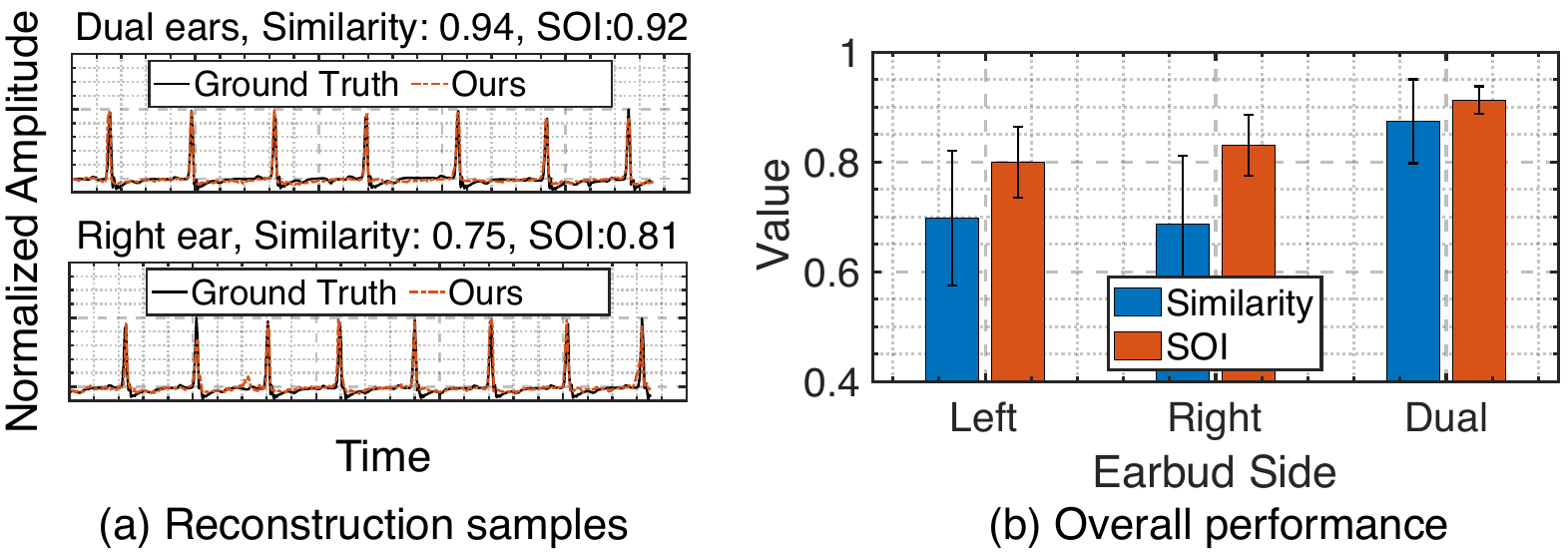}
 \caption{ECG reconstruction.}
 \label{fig:exp_ecgReconstruction}
 \vspace{-0.2in}
\end{figure}
We resampled the ECG signals from $130$ Hz to $100$ Hz to align with our model output requirements.
Figure~\ref{fig:exp_ecgReconstruction} showcases the reconstructed ECG waveform. 
A single-ear setup achieves a Similarity of $0.70$ and an SOI of $0.80$ while dual-ear setup further enhances performance, with Similarity of $0.87$ and SOI of $0.91$, respectively. This suggests \textit{\sysname's potential to reconstruct ECG signals}.

\presec
\section{Conclusion}
We have demonstrated that IMU sensors in COTS TWS earbuds can be directly leveraged for in-ear BCG sensing, enabling effective cardiac monitoring without any hardware modifications.
\sysname addresses practical challenges such as low and unreliable sampling, and motion artifacts, reconstructing SCG signals through lightweight signal processing and advanced machine learning techniques.
Our evaluation, involving 100 participants, demonstrates \sysname’s generalization across diverse BMI, age, and gender profiles.
Our case studies further highlight the transformative potential of in-ear BCG to enable a wide range of critical healthcare and HCI applications.
With its promising capabilities and real-time performance in real-world scenarios, we believe \sysname has the potential to be seamlessly integrated into future commercial products, transforming everyday earbuds into powerful, non-invasive cardiac monitors and ushering in a new era for earable cardiac monitoring.

\clearpage
\bibliographystyle{plainnat}
\bibliography{main}

\begin{thebibliography}{40}
\providecommand{\natexlab}[1]{#1}
\providecommand{\url}[1]{\texttt{#1}}
\expandafter\ifx\csname urlstyle\endcsname\relax
  \providecommand{\doi}[1]{doi: #1}\else
  \providecommand{\doi}{doi: \begingroup \urlstyle{rm}\Url}\fi

\bibitem[Apple(2024{\natexlab{a}})]{apple_CMHeadphoneMotionManager}
Apple.
\newblock Apple developer cmheadphonemotionmanager, 2024{\natexlab{a}}.
\newblock \url{https://developer.apple.com/documentation/coremotion/cmheadphonemotionmanager}.

\bibitem[Apple(2024{\natexlab{b}})]{apple_airpods}
Apple.
\newblock Apple airpods, 2024{\natexlab{b}}.
\newblock \url{https://www.apple.com/airpods/}.

\bibitem[Apple(2024{\natexlab{c}})]{apple_coreml}
Apple.
\newblock Apple developer core ml, 2024{\natexlab{c}}.
\newblock \url{https://developer.apple.com/documentation/coreml/}.

\bibitem[Apple(2024{\natexlab{d}})]{ear_tip_fit_test}
Apple.
\newblock Choose your airpods pro ear tips and use the ear tip fit test, 2024{\natexlab{d}}.
\newblock \url{https://support.apple.com/en-us/119849}.

\bibitem[Balaji et~al.(2023)Balaji, Ferlini, Kawsar, and Montanari]{balaji2023stereo}
Ananta~Narayanan Balaji, Andrea Ferlini, Fahim Kawsar, and Alessandro Montanari.
\newblock Stereo-bp: Non-invasive blood pressure sensing with earables.
\newblock In \emph{\Proc ACM HotMobile}, 2023.

\bibitem[Cao et~al.(2023)Cao, Cai, Li, Chen, and Luo]{cao2023heartprint}
Yetong Cao, Chao Cai, Fan Li, Zhe Chen, and Jun Luo.
\newblock Heartprint: Passive heart sounds authentication exploiting in-ear microphones.
\newblock In \emph{\Proc IEEE INFOCOM}, 2023.

\bibitem[Chen et~al.(2024)Chen, Yang, Fan, Guo, Xiong, and Shangguan]{chen2024exploring}
Tao Chen, Yongjie Yang, Xiaoran Fan, Xiuzhen Guo, Jie Xiong, and Longfei Shangguan.
\newblock Exploring the feasibility of remote cardiac auscultation using earphones.
\newblock In \emph{\Proc ACM MobiCom}, 2024.

\bibitem[Di~Cesare et~al.(2024)Di~Cesare, Perel, Taylor, Kabudula, Bixby, Gaziano, McGhie, Mwangi, Pervan, Narula, et~al.]{di2024heart}
Mariachiara Di~Cesare, Pablo Perel, Sean Taylor, Chodziwadziwa Kabudula, Honor Bixby, Thomas~A Gaziano, Diana~Vaca McGhie, Jeremiah Mwangi, Borjana Pervan, Jagat Narula, et~al.
\newblock The heart of the world.
\newblock \emph{Global heart}, 2024.

\bibitem[eSense(2024)]{eSense_android}
eSense.
\newblock esense android library, 2024.
\newblock \url{https://github.com/pervasive-systems/eSense-Android-Library}.

\bibitem[Fan et~al.(2021)Fan, Shangguan, Rupavatharam, Zhang, Xiong, Ma, and Howard]{fan2021headfi}
Xiaoran Fan, Longfei Shangguan, Siddharth Rupavatharam, Yanyong Zhang, Jie Xiong, Yunfei Ma, and Richard Howard.
\newblock Headfi: bringing intelligence to all headphones.
\newblock In \emph{\Proc ACM MobiCom}, 2021.

\bibitem[Fan et~al.(2023)Fan, Pearl, Howard, Shangguan, and Thormundsson]{fan2023apg}
Xiaoran Fan, David Pearl, Richard Howard, Longfei Shangguan, and Trausti Thormundsson.
\newblock Apg: Audioplethysmography for cardiac monitoring in hearables.
\newblock In \emph{\Proc ACM MobiCom}, 2023.

\bibitem[Ferlini et~al.(2021)Ferlini, Montanari, Min, Li, Sassi, and Kawsar]{ferlini2021ear}
Andrea Ferlini, Alessandro Montanari, Chulhong Min, Hongwei Li, Ugo Sassi, and Fahim Kawsar.
\newblock In-ear ppg for vital signs.
\newblock \emph{IEEE Pervasive Computing}, 2021.

\bibitem[Fu et~al.(2022)Fu, Wang, Zhong, Chen, Ren, and Zhang]{fu2022svoice}
Yongjian Fu, Shuning Wang, Linghui Zhong, Lili Chen, Ju~Ren, and Yaoxue Zhang.
\newblock Svoice: Enabling voice communication in silence via acoustic sensing on commodity devices.
\newblock In \emph{Proceedings of the 20th ACM Conference on Embedded Networked Sensor Systems}, pages 622--636, 2022.

\bibitem[Ha et~al.(2020)Ha, Assana, and Adib]{ha2020contactless}
Unsoo Ha, Salah Assana, and Fadel Adib.
\newblock Contactless seismocardiography via deep learning radars.
\newblock In \emph{\Proc ACM MobiCom}, 2020.

\bibitem[Huang et~al.(2018)Huang, Kaufmann, Aksan, Black, Hilliges, and Pons-Moll]{huang2018deep}
Yinghao Huang, Manuel Kaufmann, Emre Aksan, Michael~J Black, Otmar Hilliges, and Gerard Pons-Moll.
\newblock Deep inertial poser: Learning to reconstruct human pose from sparse inertial measurements in real time.
\newblock \emph{ACM Transactions on Graphics (TOG)}, 2018.

\bibitem[Inan et~al.(2014)Inan, Migeotte, Park, Etemadi, Tavakolian, Casanella, Zanetti, Tank, Funtova, Prisk, et~al.]{inan2014ballistocardiography}
Omer~T Inan, Pierre-Francois Migeotte, Kwang-Suk Park, Mozziyar Etemadi, Kouhyar Tavakolian, Ramon Casanella, John Zanetti, Jens Tank, Irina Funtova, G~Kim Prisk, et~al.
\newblock Ballistocardiography and seismocardiography: A review of recent advances.
\newblock \emph{IEEE journal of biomedical and health informatics}, 2014.

\bibitem[Johnston et~al.(2020)Johnston, Barrett-Jolley, Krige, and Welters]{johnston2020heart}
Brian~W Johnston, Richard Barrett-Jolley, Anton Krige, and Ingeborg~D Welters.
\newblock Heart rate variability: Measurement and emerging use in critical care medicine.
\newblock \emph{Journal of the Intensive Care Society}, 2020.

\bibitem[Kawsar et~al.(2018)Kawsar, Min, Mathur, and Montanari]{kawsar2018earables}
Fahim Kawsar, Chulhong Min, Akhil Mathur, and Alessandro Montanari.
\newblock Earables for personal-scale behavior analytics.
\newblock \emph{IEEE Pervasive Computing}, 2018.

\bibitem[Kim et~al.(2016)Kim, Ober, McMurtry, Finegan, Inan, Mukkamala, and Hahn]{kim2016ballistocardiogram}
Chang-Sei Kim, Stephanie~L Ober, M~Sean McMurtry, Barry~A Finegan, Omer~T Inan, Ramakrishna Mukkamala, and Jin-Oh Hahn.
\newblock Ballistocardiogram: Mechanism and potential for unobtrusive cardiovascular health monitoring.
\newblock \emph{Scientific reports}, 2016.

\bibitem[Lu et~al.(2013)Lu, Tsao, Matsuda, and Hori]{lu2013speech}
Xugang Lu, Yu~Tsao, Shigeki Matsuda, and Chiori Hori.
\newblock Speech enhancement based on deep denoising autoencoder.
\newblock In \emph{Interspeech}, 2013.

\bibitem[Mahmood et~al.(2019)Mahmood, Ghorbani, Troje, Pons-Moll, and Black]{mahmood2019amass}
Naureen Mahmood, Nima Ghorbani, Nikolaus~F Troje, Gerard Pons-Moll, and Michael~J Black.
\newblock Amass: Archive of motion capture as surface shapes.
\newblock In \emph{\Proc ICCV}, 2019.

\bibitem[McKay(1970)]{mckay1970signal}
George~A McKay.
\newblock Signal-to-noise ratio in the bandpass output of a discriminator.
\newblock \emph{IEEE Transactions on Aerospace and Electronic Systems}, 1970.

\bibitem[Mukaka(2012)]{mukaka2012guide}
Mavuto~M Mukaka.
\newblock A guide to appropriate use of correlation coefficient in medical research.
\newblock \emph{Malawi medical journal}, 24\penalty0 (3):\penalty0 69--71, 2012.

\bibitem[Nagai et~al.(2017)Nagai, Anzai, and Wang]{nagai2017motion}
Shuto Nagai, Daisuke Anzai, and Jianqing Wang.
\newblock Motion artefact removals for wearable ecg using stationary wavelet transform.
\newblock \emph{Healthcare technology letters}, 2017.

\bibitem[Nelson and Allen(2019)]{nelson2019accuracy}
Benjamin~W Nelson and Nicholas~B Allen.
\newblock Accuracy of consumer wearable heart rate measurement during an ecologically valid 24-hour period: intraindividual validation study.
\newblock \emph{JMIR mHealth and uHealth}, 2019.

\bibitem[Nokia Bell~Labs(2019)]{eSense_doc}
Cambridge Nokia Bell~Labs.
\newblock esense user documentation, 2019.
\newblock \url{https://www.esense.io/share/eSense-User-Documentation.pdf}.

\bibitem[Ochoa and Revilla(2022)]{ochoa2022signal}
Emilio~J Ochoa and Luis~C Revilla.
\newblock Signal filtering and peak analysis of ballistocardiography for heartbeat detection.
\newblock In \emph{International Conference on Biomedical and Health Informatics}. Springer, 2022.

\bibitem[Polor(2024)]{polor10_usermanual}
Polor.
\newblock Polor h10 user manual, 2024.
\newblock \url{https://support.polar.com/e_manuals/h10-heart-rate-sensor/polar-h10-user-manual-english/manual.pdf}.

\bibitem[ResearchAndMarkets.com(2023)]{TWSmarket}
ResearchAndMarkets.com.
\newblock Global true wireless stereo earbuds market, 2022-2030: Health meets entertainment - the fusion of hearing aid functionalities in tws earbuds.
\newblock \href{https://www.globenewswire.com/en/news-release/2023/10/20/2763898/28124/en/Global-True-Wireless-Stereo-Earbuds-Market-2022-2030-Health-Meets-Entertainment-The-Fusion-of-Hearing-Aid-Functionalities-in-TWS-Earbuds.html}{Global TWS Market Report}, 2023.

\bibitem[Romero et~al.(2024)Romero, Ferlini, Spathis, Dang, Farrahi, Kawsar, and Montanari]{romero2024optibreathe}
Julia Romero, Andrea Ferlini, Dimitris Spathis, Ting Dang, Katayoun Farrahi, Fahim Kawsar, and Alessandro Montanari.
\newblock Optibreathe: An earable-based ppg system for continuous respiration rate, breathing phase, and tidal volume monitoring.
\newblock In \emph{\Proc ACM HotMobile}, 2024.

\bibitem[Sadek et~al.(2019)Sadek, Biswas, and Abdulrazak]{sadek2019ballistocardiogram}
Ibrahim Sadek, Jit Biswas, and Bessam Abdulrazak.
\newblock Ballistocardiogram signal processing: A review.
\newblock \emph{Health information science and systems}, 2019.

\bibitem[Seok et~al.(2023)Seok, Song, An, and Lee]{seok2023photoplethysmogram}
Chae~Lin Seok, Young~Do Song, Byeong~Seon An, and Eui~Chul Lee.
\newblock Photoplethysmogram biometric authentication using a 1d siamese network.
\newblock \emph{Sensors}, 2023.

\bibitem[Seok et~al.(2021)Seok, Lee, Kim, Cho, and Kim]{seok2021motion}
Dongyeol Seok, Sanghyun Lee, Minjae Kim, Jaeouk Cho, and Chul Kim.
\newblock Motion artifact removal techniques for wearable eeg and ppg sensor systems.
\newblock \emph{Frontiers in Electronics}, 2021.

\bibitem[Sherman(2024)]{Apple_HearingAid}
Lauren Sherman.
\newblock Airpods as hearing aids: New features announced, 2024.
\newblock \url{https://www.ncoa.org/adviser/hearing-aids/airpods-hearing-aids/}.

\bibitem[Tipparaju et~al.(2021)Tipparaju, Mallires, Wang, Tsow, and Xian]{tipparaju2021mitigation}
Vishal~Varun Tipparaju, Kyle~R Mallires, Di~Wang, Francis Tsow, and Xiaojun Xian.
\newblock Mitigation of data packet loss in bluetooth low energy-based wearable healthcare ecosystem.
\newblock \emph{Biosensors}, 2021.

\bibitem[Wang et~al.(2023)Wang, Wang, Zhang, Ma, Zhang, Dai, Xu, Li, and Gu]{wang2023knowing}
Lei Wang, Xingwei Wang, Dalin Zhang, Xiaolei Ma, Yong Zhang, Haipeng Dai, Chenren Xu, Zhijun Li, and Tao Gu.
\newblock Knowing your heart condition anytime: User-independent ecg measurement using commercial mobile phones.
\newblock \emph{\Proc ACM IMWUT (UbiComp)}, 2023.

\bibitem[Winokur et~al.(2012)Winokur, Da~He, and Sodini]{winokur2012wearable}
Eric~S Winokur, David Da~He, and Charles~G Sodini.
\newblock A wearable vital signs monitor at the ear for continuous heart rate and pulse transit time measurements.
\newblock In \emph{2012 Annual International Conference of the IEEE Engineering in Medicine and Biology Society}. IEEE, 2012.

\bibitem[Zhang et~al.(2023{\natexlab{a}})Zhang, Shi, Walker, Ye, Wang, Saxena, and Chen]{zhang2023passive}
Tianfang Zhang, Cong Shi, Payton Walker, Zhengkun Ye, Yan Wang, Nitesh Saxena, and Yingying Chen.
\newblock Passive vital sign monitoring via facial vibrations leveraging ar/vr headsets.
\newblock In \emph{\Proc ACM MobiSys}, 2023{\natexlab{a}}.

\bibitem[Zhang et~al.(2023{\natexlab{b}})Zhang, Ye, Mahdad, Akanda, Shi, Wang, Saxena, and Chen]{zhang2023facereader}
Tianfang Zhang, Zhengkun Ye, Ahmed~Tanvir Mahdad, Md~Mojibur Rahman~Redoy Akanda, Cong Shi, Yan Wang, Nitesh Saxena, and Yingying Chen.
\newblock Facereader: Unobtrusively mining vital signs and vital sign embedded sensitive info via ar/vr motion sensors.
\newblock In \emph{\Proc ACM CCS}, 2023{\natexlab{b}}.

\bibitem[Zhao et~al.(2024)Zhao, Li, Xie, Xie, Zhang, Zhang, and Wang]{zhao2024hearbp}
Zhiyuan Zhao, Fan Li, Yadong Xie, Huanran Xie, Kerui Zhang, Li~Zhang, and Yu~Wang.
\newblock Hearbp: Hear your blood pressure via in-ear acoustic sensing based on heart sounds.
\newblock In \emph{\Proc IEEE INFOCOM}, 2024.

\end{thebibliography}

\appendix

\end{document}